\documentclass{aa}

\usepackage{graphicx,array,txfonts}
\usepackage{natbib}
\usepackage{nicefrac}
\usepackage{hyperref}
\usepackage{tikz}
\usepackage{stmaryrd}

\bibpunct{(}{)}{;}{a}{}{,}

\newcolumntype{L}[1]{>{\raggedright\let\newline\\\arraybackslash\hspace{0pt}}m{#1}}
\newcolumntype{C}[1]{>{\centering  \let\newline\\\arraybackslash\hspace{0pt}}m{#1}}

\usetikzlibrary{arrows,calc,fadings,decorations.pathreplacing,positioning,shapes,shadows,intersections,backgrounds,quotes,angles}
\tikzstyle{split_cell} = [rectangle, draw=black, anchor=north, rectangle split, rectangle split parts=2, align=center, text width=6cm]
\tikzstyle{myarrow}=[-latex', >=open triangle 90, thick]
\tikzstyle{line}=[-, thick]

\hypersetup{
    colorlinks=true,
    linkcolor=blue,
    citecolor=blue,
    filecolor=magenta,
    urlcolor=cyan,
    }

\def \dd{\mathrm{d}}
\def \ee{\mathrm{e}}
\def \ss{\mathrm{s}}
\def \mB{{\mathcal{B}}}
\def \mC{{\mathcal{C}}}
\def \mF{{\mathcal{F}}}
\def \mG{{\mathcal{G}}}
\def \mK{{\mathcal{K}}}
\def \mT{{\mathcal{T}}}
\def \bze{{\bf 0}}
\def \bE{{\bf E}}
\def \bg{{\bf g}}
\def \by{{\bf y}}
\def \eos{EoS\xspace}
\def \eoss{EoSs\xspace}
\def \teff{T_{\rm eff}\xspace}
\def \ttau{T(\tau)\xspace}
\def \logg{\log g\xspace}
\def \amlt{\alpha_{\rm MLT}\xspace}
\def \feh{{\rm [Fe/H]}}
\def \cesam{CESAM\xspace}
\def \cesamk{CESAM2k\xspace}
\def \cestam{CESTAM\xspace}
\def \cesamxx{Cesam2k20\xspace}
\def \corot{CoRoT\@\xspace}
\def \plato{\textsc{plato}\@\xspace}

\def \ds{\displaystyle}
\def \f{\frac}
\def \nf{\nicefrac}
\newcommand{\std}[1]{\left.{#1}\right|}
\newcommand{\stp}[1]{\left({#1}\right)}
\newcommand{\stb}[1]{\left[{#1}\right]}
\newcommand{\stbb}[1]{\left\llbracket{#1}\right\rrbracket}
\newcommand{\pd}[2]{\frac{\partial #1}{\partial #2}}
\newcommand{\lpd}[2]{{\partial #1}/{\partial #2}}
\newcommand{\ldiff}[2]{{\dd #1}/{\dd #2}}
\newcommand{\diff}[2]{\frac{\dd #1}{\dd #2}}
\newcommand{\tdiff}[3]{\frac{\dd^{#3} #1}{\dd #2^{#3}}}

\makeatletter
        \newcommand{\vast}{\bBigg@{3}}
\makeatother

\def \none{\multicolumn{1}{c}{--}}

\begin{document}

    \title{Cesam2k20: A code for a new generation of stellar evolution models}
    \titlerunning{Cesam2k20: A code for a new generation of stellar evolution models}
    \subtitle{I. Description of the code}

    \author{
        L.~Manchon\inst{1}
        \and M.~Deal\inst{2}
        \and J.~P.~C.~Marques \inst{3}
        \and Y.~Lebreton\inst{1,4}}
    \institute{
    LIRA, Observatoire de Paris, Université PSL, Sorbonne Université, Université Paris Cité, CY Cergy Paris Université, CNRS, 92190 Meudon, France\\
    \texttt{email: louis.manchon@obspm.fr}
    \and LUPM, Universit\'e de Montpellier, CNRS, place Eug\`ene Bataillon, 34095 Montpellier, France
    \and Institut d'Astrophysique Spatiale, Université Paris-Saclay, Orsay, France
    \and Universit\'e de Rennes, CNRS, IPR (Institut de Physique de Rennes) -- UMR 6251, 35000 Rennes, France}

    \date{Received XXX / Accepted XXX}

    \abstract{We present Cesam2k20, the latest version of the hydrostatic stellar evolution code CESAM originally developed by P. Morel and collaborators. Over the last three decades, it has undergone many improvements and has been extensively tested against other stellar evolution codes before being selected to compute the first-generation grid of stellar models for the PLATO mission. Among all the developments made thus far, Cesam2k20 now implements state-of-the-art models for the transport of chemical elements and angular momentum. It was recently  made publicly available with an ecosystem of other codes interfaced with it:  1D and 2D oscillation codes ADIPLS and ACOR,  optimisation program OSM, and Python utility package \texttt{pycesam}. This paper recalls the numerical peculiarities of Cesam2k20, namely, the use of a collocation method where the structure variables are decomposed as piecewise polynomials projected on a B-spline basis. Here, we review the options available for modelling the different physical processes. In particular, we illustrate the improvements made in the transport of chemical elements and angular momentum with a series of standard and non-standard solar models.}


    \maketitle

    \section{Introduction}

    \cesamxx is a hydrostatic stellar evolution code that comprises the latest version of the Code d'Évolution Stellaire, Adaptatif et Modulaire (\cesam) stellar evolution code originally developed by Pierre Morel and collaborators. This code was first used to produce standard solar models \citep{Berthomieu1993}, with the numerical details presented in \citet[hereafter M97]{Morel1997}. \cesam was originally written in FORTRAN 77 for its first four versions and later translated into Fortran 90 for the fifth. In the early 2000s, \cesam was renamed \cesamk. \citet[hereafter ML08]{Morel2008} published a full description of this new version as part of the scientific preparation for the Convection Rotation and Transits mission (CoRoT; \citealt{Baglin2006}).

    The availability of precise asteroseismic data on the stellar interior and the first measurements of internal stellar rotation have motivated sizable efforts to test models of angular momentum transport processes. Such models were implemented in \cesamk, then renamed \cestam (where T stands for Transport). Using this code, \citet[hereafter M13]{Marques2013} and \citet{Goupil2013a} showed that our understanding of angular momentum transport is far from  complete \citep[see][for a review]{Goupil2013b,Aerts2019}. A lot of work has also been put into improving the user-friendliness of the code, especially with respect to the automated installation process and new Python (graphical) interface.

    The next developments, results of intense collaborative efforts, have been important in the following years. They are now gathered in a new version of the \cesam code for the 2020 decade: \cesamxx\footnote{\url{https://www.ias.u-psud.fr/cesam2k20/}}. The repository\footnote{\url{https://git.ias.u-psud.fr/joao.marques/cesam2k20_releases}} hosting the public versions will be regularly updated with new versions of \cesamxx.

    This state-of-the-art version has benefited from successive validations and comparisons against other stellar evolution codes. The first validation dates back to the year that followed the publication of the seminal paper with a comparison by \citet{Turck-Chieze1998} of two versions of CESAM with earlier versions or predecessors of ASTEC \citep{Christensen-Dalsgaard1982,Christensen-Dalsgaard2008b}, GARSTEC \citep{Schlattl1997,Weiss2008}, CLES \citep{Scuflaire2008}, and the Montreal code \citep{Turcotte1998,Richer2000}. Over the next few years, it was tested on Hyades binaries \citep{Lastennet1999}, the Hyades cluster \citep{deBruijne2001}, and red stragglers \citep{Fernandes2004}, as well as against the Geneva \citep{Eggenberger2008} and Padova/PARSEC \citep{Bressan1993,Bressan2012} codes. Prior to the launch of the CoRoT mission, the  Evolution and Seismic Tools Activity (ESTA) model comparison project \citep{Lebreton2008a, Lebreton2008b, Montalban2008, lebreton2008c, Marconi2008,Goupil2008}, extensively tested almost all the stellar evolution codes existing at that time \citep[for a description of the codes included in the ESTA comparison, see][]{Monteiro2009}. Similar comparisons were performed for higher mass stars \citep{Weiss2007} and later for the special case of red giant modelling \citep{SilvaAguirre2020,Christensen-Dalsgaard2020}. The code has been part of many stellar model comparisons in the context the CoRoT mission with CoRoT targets \citep{Castro2021} against the TGEC code \citep{HuiBonHoa2008} and of those the Kepler mission \citep{SilvaAguirre2013, Reese2016, SilvaAguirre2017}. More recently, the treatment of atomic diffusion \citep{Campilho2022} and of the overshoot \citep{Noll2023} have been compared with MESA \citep{Paxton2013}.

    This code or previous versions have been used to model a wide variety of stars. Of course, starting with the Sun \citep{Berthomieu1993,Brun1998,Brun1999,Samadi2001,Couvidat2003,Mecheri2004} and including solar-like oscillators and Sun-like stars have been primary targets of \cesamxx users, including \cite{Piau2002,Ballot2004,Deheuvels2010a,Deheuvels2010b,Metcalfe2010}.  However, it has also been used for modelling magnetic stars \citep{Cunha2006,Alecian2007a,Alecian2007b,Alecian2009, Villebrun2019, AlecianE2020}, binary stars $\iota$ Pegasi A\&B \citep{Morel2000a}, $\alpha$ Centauri A \& B \citep{Morel2000b,Thevenin2002,Kervella2003}, 85 Peg \citep{Fernandes2002}, $\zeta$ Her A \& B \citep{Lebreton1993,Morel2001}, EK Cephei A \& B \citep{Marques2004}, 61 Cyg A \& B \citep{Kervella2008}, and samples of nearby binaries or Gaia and Kepler targets \citep{Fernandes1998, Appourchaux2015,Kiefer2016,Kiefer2018, Halbwachs2020}, as well as ensembles of stars or cluster members \citep{Fernandes1996, Cayrel1999, Perryman1998, Lebreton1999, Cayrel2000, Lebreton2001, Torres2011} and various classes of pulsators such as solar-like pulsators; these include exoplanet hosts, pre-main sequence stars,  subgiants, and red giants \citep{Suran2001, Lebreton2012, Baudin2012, Lebreton2014, Crida2018, Ligi2019, Huber2019, Chaplin2020}, $\lambda$ Bootis stars \citep{Paunzen1998}, $\delta$ Scuti stars \citep{Hernandez1998,Michel1999,Samadi2002, Suarez2002,Amado2004,Chapellier2004,Suarez2005a,Poretti2005,FoxMachado2006,Suarez2006,Casas2006,Suarez2007,GarciaHernandez2009,Poretti2009}, $\gamma$ Doradus stars \citep{Moya2005,Suarez2005b,Ouazzani2019}, blue supergiants \citep{Godart2009, Lebreton2009}, and Cepheids \citep{Cordier2002,Cordier2003,Morel2010a}. Finally, \cesamxx models have supported theoretical works, improving our understanding of asteroseismology and deriving seismic diagnostics \citep{Dintrans2000,Mazumdar2001,Samadi2001,Samadi2003,Samadi2006,Belkacem2008,Belkacem2009,Belkacem2015a,Belkacem2015b, Deheuvels2016, Houdayer2021} to exploring dark matter models \citep{Lopes2002,Lopes2003,Lopes2011,Casanellas2011, Casanellas2015}. At the moment, \cesamxx can compute stellar models for mass below $20M_\odot$ from the PMS to He-burning; however, low-mass star models cannot pass the He flash. For models to pass the flash, the evolution must be stopped during the red giant branch (RGB) above the red giant bump.

    This paper is aimed at presenting the new options available in \cesamxx\ (rather than performing validation, tests, or comparisons). Such developments can be found  in the above references. Section \ref{sect:cesam} provides a concise description of \cesamxx, as the numerics are already detailed in depth in M97 and ML08. Newly available equations of state (\eos), opacity and nuclear reaction tables are summarized in Sect. \ref{sect:eos_opac}. Developments in the treatment of convection, evolution of chemicals and of angular momentum, and the restitution of the atmosphere are  detailed in Sect. \ref{sect:conv}, Sect. \ref{sect:chim_evol}, and Sects. \ref{sect:amt} and \ref{sect:atmosphere}, respectively. As \cesamxx is moving forward to be able to compute advanced evolutionary phases, Sect. \ref{sect:mass_loss} describes the new mass loss prescriptions implemented. Finally, we illustrate what \cesamxx is capable of with standard and non-standard solar models in Sect \ref{sect:nssm}.

    \section{Cesam2k20's numerics in brief}
    \label{sect:cesam}

    Most stellar evolution codes rely on finite element schemes to solve the stellar structure equations and the transport of chemicals. With these schemes, the star is divided into layers and $k$-th order derivatives are approximated using the values of each quantity at the faces of the nearest layers. Those methods are fast, well tested and easy to implement. However, they provide solutions only at the faces of each layer and not between them. Furthermore, during stellar evolution, the mesh frequently needs to be adapted to resolve regions with strong gradients, and it can be useful to adopt different meshes for solving different systems of equations (i.e. a mesh dedicated to structure equations, another one for the transport of chemicals, etc.). The re-meshing is hindered by the fact that the solutions are only known at a few discrete points.

    \cesamxx follows quite a peculiar path. Solutions are represented by a linear combination of B-Splines \citep{Schumaker2007}. The B-Splines form an orthogonal basis of piecewise continuous polynomials. The representation of the solution to an equation as a linear combination of B-Splines provides an (approximate) knowledge of this quantity everywhere, by only knowing the exact solution at particular points, known as collocation points. Another advantage is that a different mesh for different systems of equations can be used. For instance, the masses at which the structure equations are solved are not the same as the ones that the equations for the transport of chemicals are solved for. This could also be the case for the transport of angular momentum.

    \subsection{Grid point allocation}

    Each spherical shell in the model is labelled with its radius, $r_i$, and the mass, $m_i$, enclosed in it. We need to increase the resolution where the gradients of the significant quantities are the strongest. Therefore, the layers should be located in a way that minimizes these gradients or at least that keeps them below a certain threshold. To that end, we introduced a quantity, $Q$, which serves as the spacing function. This quantity is defined so that, at a given time step, the variation of $Q$ between two consecutive layers of mass, $m_i$ and $m_{i+1}$, is constant, expressed as
    \begin{eqnarray}
        Q(m_{i+1}, t) - Q(m_i) = \std{\diff{Q}{q}}_t \equiv \psi(t) = {\rm cst,}
    \end{eqnarray}
    and with
    \begin{eqnarray}
        \std{\tdiff{Q}{q}{2}}_t = 0,
    \end{eqnarray}
    where $q$ is called the index function and takes integer values at each layer, from $1$ to $n$, with $n$ the total number of layers. The function $\psi$ can be written as a function of the mass,
    \begin{equation}
        \std{\diff{Q}{q}}_t = \std{\diff{Q}{m}}_t \std{\diff{m}{q}}_t =\theta(t) \std{\diff{m}{q}}_t = \psi(t),
    \end{equation}
    where $\theta(t)$ has been identified as $\std{\ldiff{Q}{m}}_t$. We end up with two more differential equations, which are to be added to the four structure equations,
    \begin{eqnarray}
        \diff{m}{q} = \f{\psi}{\theta} & \textrm{and} & \diff{\psi}{q} = 0.
        \label{eq:dmdq}
    \end{eqnarray}
    Equations \eqref{eq:dmdq} are complemented by boundary conditions: at $q = 1$, $m = 0$ and at $q = n$, $m = M_\star$.

    We still need a definition for the spacing function $Q$, that must be a class $\mC^2$ function, strictly increasing. A simple choice is
    \begin{equation}
        Q(m, t) = \f{p}{\Delta p} + \f{T}{\Delta T} + \f{L}{\Delta L} + \f{r}{\Delta r} + \f{m}{\Delta m},
        \label{eq:Q_general}
    \end{equation}
    where $p$ is the pressure, $T$ the temperature, and $L$ the luminosity, while the terms $\Delta f$ represent the repartition factors and  are aimed at weighting the importance of each quantity. With this definition, $Q(m_{i+1}, t) - Q(m_k)$ behaves as the sum of the normalized gradients of each independent variable. Each weight $\Delta f$ can be chosen when running a \cesamxx model. However, years of use of \cesamxx and preceding version have shown that $(1/\Delta p; 1/\Delta T; 1/\Delta L; 1/\Delta r; 1/\Delta m) = (-1, -1, 0, 0, 15)$ is a very good choice, at least before He burning phases.

    \subsection{Dimensionless structure equations}
    \label{subsection:dimensionless_cestam}

    \cesamxx assumes hydrostatic equilibrium and solves six differential equations associated with six independent variables $r, p, T, L, m,$ and $\psi$. The equations for $r, p, T, L$ are the usual structure expressions (and they can be found e.g. in M97, without rotation), while the expressions for $\psi$ and $\theta$ are given in Eq. \eqref{eq:dmdq}.  To improve floating-point precision, \cesamxx actually solves equation for dimensionless variables: $\xi = \ln p$, $\eta = \ln T$,  $\lambda = (L/L_\odot)^{a}$, $\zeta = (r/R_\odot)^2$, $\nu = (m / M_\odot)^{2/3}$, and $\psi$, which is already dimensionless. The power, $a$, on the luminosity is either $2/3$ or $1$. In the past, $a=2/3$ was used to avoid derivability issues near the centre, but it is now possible to use $a=1$. In terms of these dimensionless variables, the structure equation system is expressed as
    \begin{equation}
        \left\{
        \begin{array}{l}
        \ds \pd{\xi}{q}     = \ds \f{-3\mG}{8\pi}\stp{\f{M_\odot}{R_\odot^2}}^2 \stp{\f{\nu}{\zeta}}^2 \exp(-\xi)\f{\psi}{\theta}, \\[10pt]
        \ds \pd{\eta}{q}    = \ds \pd{\xi}{q} \nabla, \\[10pt]
        \ds \pd{\zeta}{q}   = \ds \f{3}{4\pi}\f{M_\odot}{R_\odot^3}\stp{\f{\nu}{\zeta}}^{\nf{1}{2}}\f{1}{\rho}\f{\psi}{\theta}, \\[10pt]
        \ds \pd{\lambda}{q} = \f{M_\odot \nu^{\nf{1}{2 }}}{L_\odot \lambda^{\nf{1}{2}}}\Lambda\f{\psi}{\theta},~\textrm{with}~a = 1, \\[10pt]
        \ds \pd{\lambda}{q} = \f{3}{2}\f{M_\odot \nu^{\nf{1 }{2}}}{L_\odot}\Lambda\f{\psi}{\theta},~\textrm{with}~a = \f{2}{3}, \\[10pt]
        \ds \pd{\nu}{q}     = \ds \f{\psi}{\theta}, \\[10pt]
        \ds \pd{\psi}{q}    = 0,
        \end{array}
        \right.
    \label{eq:cestam_struct_eqs}
    \end{equation}
    where $\mG$ is the gravitational constant, $\nabla \equiv \ldiff{\ln T}{\ln p}$ is the temperature gradient, $\rho$ is the density, and $M_\odot$, $R_\odot$, and $L_\odot$ are  the solar mass, radius, and luminosity, respectively.

    \subsection{Collocation method}
    \label{subsection:colloc}

    Let us rewrite the above system of first order ordinary differential equations in a more compact way. By denoting $\by = (y_1, y_2, y_3, y_4, y_5, y_6) = (\xi, \eta, \zeta, \lambda, \nu, \psi)$ the unknowns, the system in Eq. \eqref{eq:cestam_struct_eqs} may be written as    \begin{eqnarray}
        \bE(q; \by, \by') = \diff{\by}{q} - \bg(q, \by) = \bze &\textrm{with} & q \in [q_1, q_n],
        \label{eq:system_E_compact}
    \end{eqnarray}
    with $\bg$ a suitable vector of functions representing the differential system.

    The system in Eq. \eqref{eq:cestam_struct_eqs}, while keeping only $a = 1$ for simplicity, is then
    \begin{align}
        \left\{
        \begin{array}{l}
            \ds E_1 = 0 = \ds \pd{y_1}{q} - \f{-3\mG}{8\pi}\stp{\f{M_\odot}{R_\odot^2}}^2 \stp{\f{y_5}{y_3}}^2 \exp(-y_1)\f{y_6}{\theta} , \\[10pt]
            \ds E_2 = 0 = \ds \pd{y_2}{q} - \pd{y_1}{q} \nabla, \\[10pt]
            \ds E_3 = 0 = \ds \pd{y_3}{q} - \f{3}{4\pi}\f{M_\odot}{R_\odot^3}\stp{\f{y_5}{y_3}}^{\nf{1}{2}}\f{1}{\rho}\f{y_6}{\theta} , \\[10pt]
            \ds E_4 = 0 = \ds \pd{y_4}{q} - \f{M_\odot y_5^{\nf{1}{2}}}{L_\odot y_4^{\nf{1}{2}}}\Lambda\f{y_6}{\theta}, \\[10pt]
            \ds E_5 = 0 = \ds \pd{y_5}{q} - \f{y_6}{\theta} , \\[10pt]
            \ds E_6 = 0 = \ds \pd{y_6}{q}.
        \end{array}
        \right.
        \label{eq:system_E_struct}
    \end{align}
    This system is also supplemented with a set of bottom and top boundary conditions, $\bE^{\rm b}(q_1, \by)$ and $\bE^{\rm t}(q_n, \by)$. The solution is the set of functions $\by(q)$ that makes  $E_k(q) = 0, \forall k\in \llbracket 1;6\rrbracket$.

    This system is solved using a pseudo-spectral method called the collocation method \citep[e.g.][]{DeBoor2001}. Unknown functions  $\{y_i(q)\}_{i = 1}^6$ will be decomposed as a linear combination of B-Splines. We let $\{q_i\}_{i=0}^n$ be a set of points that verify the condition $a = q_0 < q_1\cdots < q_n = b$, with $a$ and $b$ as the limits of the interval in which Eq. \eqref{eq:system_E_compact} is to be solved. We denote $\{N_j^m\}_{j=1}^M$ the basis of B-Splines of the vector space of dimension $M$ of all the piecewise polynomials (of the order\footnote{We recall that the order of a polynomial is its degree minus unity.} $m$) that match at $\{q_i\}_{i=1}^{n-1}$. Any unknown $y_k$ of Eq. \eqref{eq:system_E_compact} can be decomposed as
    \begin{equation}
        y_k(q)= \sum_{j = 1}^M y_{kj} N_j^m(q),
    \end{equation}
    and
    \begin{equation}
        y'_k(q) = \diff{y_k}{q} = \sum_{j = 1}^M y_{kj} \diff{N_j^m}{q}.
    \end{equation}
    Finally, we define $\mK = \llbracket1;6\rrbracket$ and $\mB \subseteq \mK$ (resp. $\mT \subseteq \mK$), which is the set of indices of the unknowns for which a bottom (resp. top) boundary condition is provided. We end up with a set of equations taking the  following form:\ 
    \begin{itemize}
    \item At the bottom: $E^{\rm b}_k$, with $k\in \mB$ is
    \begin{equation}
        E^{\rm b}_k\stp{\sum_{j = 1}^M y_{1,j} N_j^m(q_1); \ldots; \sum_{j = 1}^M y_{6,j} N_j^m(q_1) } = 0.
    \end{equation}
    \item At the top: $E^{\rm t}_k$, with $k\in \mT$ is
    \begin{equation}
        E^{\rm t}_k\stp{\sum_{j = 1}^M y_{1,j} N_j^m(q_n); \ldots; \sum_{j = 1}^M y_{6,j} N_j^m(q_n) } = 0.
    \end{equation}
    \item Elsewhere ($q\in [q_1; q_n]$), for $k \in \mK$,
    \begin{eqnarray}
        E_k\vast(q; \sum_{j = 1}^M y_{1,j} N_j^m; \ldots; \sum_{j = 1}^M y_{6,j} N_j^m; & \nonumber\\
                \sum_{j = 1}^M y_{1,j} \diff{N_j^m}{q}; \ldots; \sum_{j = 1}^M y_{6,j} \diff{N_j^m}{q}\vast) &= 0.
    \end{eqnarray}\\
    \end{itemize}
    The coefficients $y_{k,j}$ become the unknowns of the system.

    On $M-1$ collocation points $c_k\in ]q_1, q_n[$, chosen using Gauss-Legendre quadrature \citep[see][for details]{DeBoor2001}, the coefficients $y_{i,j}$ are found using an iterative method: the     Newton-Raphson method. At first, $y_{k,j}$ are initialized with the solution found at the previous time step. Then, at a given iteration $p\geq0$, it is estimated by 
    \begin{itemize}
    \item At the bottom: $\forall i\in \mB$,
        \begin{eqnarray}
            &\ds E_i^{\rm b}\stp{\sum_{j = 1}^M y_{1,j}^p N_j^m(q_1); \ldots; \sum_{j = 1}^M y_{n_\ee,j}^p N_j^m(q_1) } \nonumber\\
            &\ds = \sum_{l=1}^{n_\ee} \sum_{j=1}^{M} \pd{E_i^{\rm b}}{y_l}N_j^m(q_1) \dd y_{lj}^p.
        \end{eqnarray}
    \item At the top: $\forall i\in \mT$,
we have        \begin{eqnarray}
            &\ds E_i^{\rm t}\stp{\sum_{j = 1}^M y_{1,j}^p N_j^m(q_n); \ldots; \sum_{j = 1}^M y_{n_\ee,j}^p N_j^m(q_n) } \nonumber\\
            &\ds = \sum_{l=1}^{n_\ee} \sum_{j=1}^{M} \pd{E_i^{\rm t}}{y_l}N_j^m(q_n) \dd y_{lj}^p.
        \end{eqnarray}
    \item Elsewhere, $\forall j \in \mK$, $\forall k\in \llbracket1;M-1\rrbracket$,
we have        \begin{eqnarray}
            & \ds E_i\vast(c_k; \sum_{j = 1}^M y_{1,j}^p N_j^m(c_k); \ldots; \sum_{j = 1}^M y_{n_\ee,j}^p N_j^m(c_k); \nonumber\\
            & \ds \sum_{j = 1}^M y_{1,j}^p \diff{N_j^m}{q}(c_k); \ldots; \sum_{j = 1}^M y_{n_\ee,j}^p \diff{N_j^m}{q}(c _k) \vast) \nonumber\\
            & \ds = \sum_{l=1}^{n_\ee} \sum_{j=1}^{M} \stp{\pd{E_i}{y_l}N_j^m(c_k) + \pd{E_i}{y_l'} \diff{N_j^m}{q}(c_k)} \dd y_{lj}^p.
        \end{eqnarray}
    \end{itemize}

    The quantities $\dd y_{lj}^p$ are small corrections to the coefficients $y_{lj}^p$ and are the unknowns of the Newton-Raphson scheme. With the value of $\dd y_{lj}^p$, we can determine the value of $y_{lj}$ at next iteration:
    \begin{equation}
        y_{lj}^{p+1} = y_{lj}^p - \dd y_{lj}^p; \quad \forall l \in \stbb{1;n_\ee}, \forall j \in \stbb{1;M}.
    \end{equation}
    The collocation method offers the advantage of superconvergence. Indeed, instead of reaching a precision of order $m$ ($m$ is the order of the B-Splines), the collocation method, by choosing the point in a judicious way, reaches a  precision of the order of $2m$.

    \subsection{General flowchart of Cesam2k20}
    \label{subsection:cestam_flowchart}

    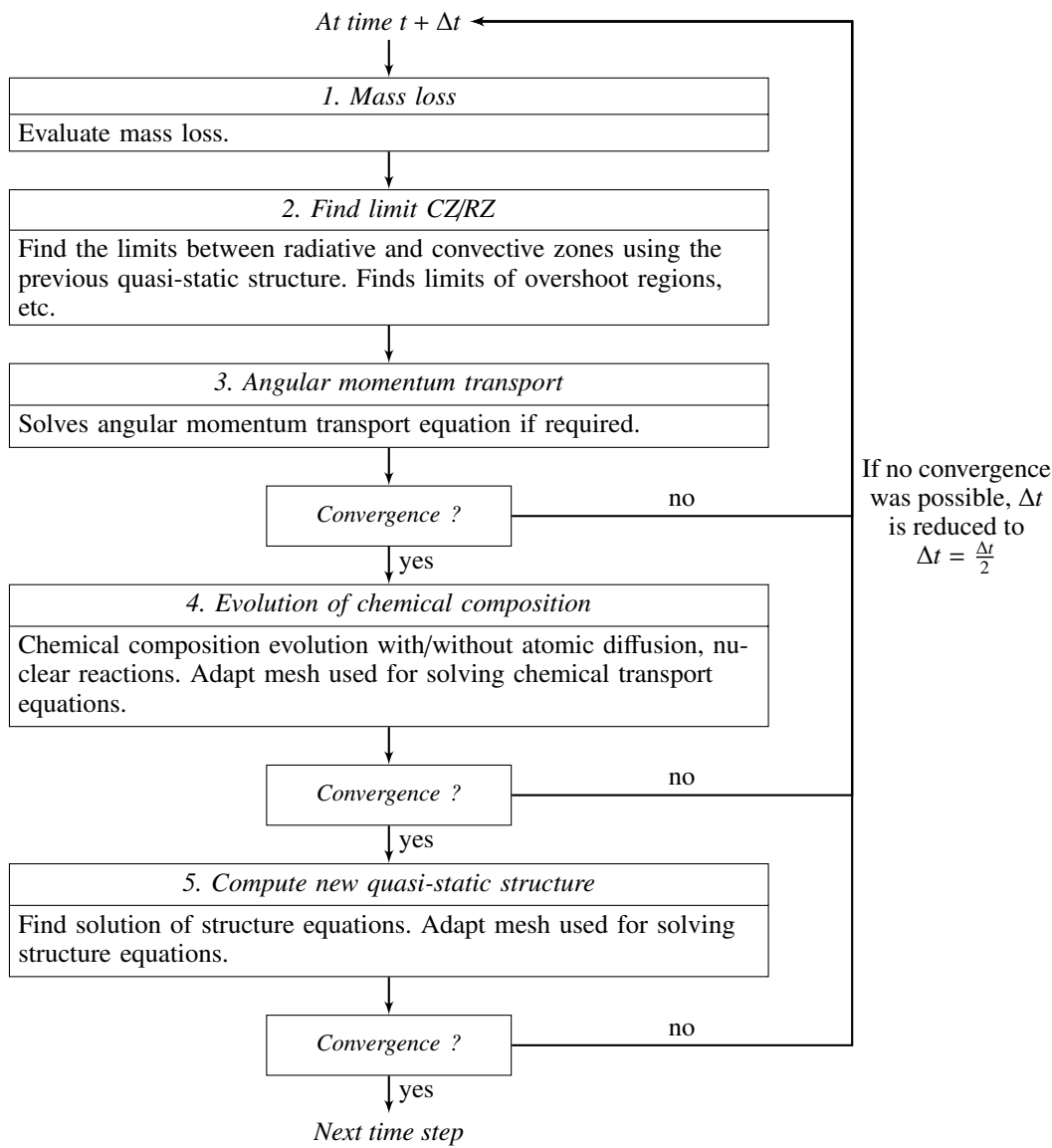
\begin{figure*}
    \begin{center}
    \begin{tikzpicture}
        \node (input) [align=center, node distance=1mm] {\textit{At time $t+\Delta t$}};

        \node (init) [split_cell, below=of input, text width=9.8cm, yshift=0.5cm]
            {\textit{1. Mass loss}
            \nodepart[align=left]{second} Evaluate mass loss.};

        \node (limzc) [split_cell, below=of init, text width=9.8cm, yshift=0.5cm]
            {\textit{2. Find limit CZ/RZ}
            \nodepart[align=left]{second} Find the limits between radiative and convective zones using the previous quasi-static structure. Finds limits of overshoot regions, etc.};

        \node (rot) [split_cell, below=of limzc, text width=9.8cm, yshift=0.5cm]
            {\textit{3. Angular momentum transport}
            \nodepart[align=left]{second} Solves angular momentum transport equation if required.};

        \node (converge_1) [rectangle, draw=black, anchor=north, yshift=0.5cm, align=center, text width=3cm, minimum size=8mm, below=of rot] {\small \textit{Convergence ?}};

        \node (chim) [split_cell, below=of converge_1, text width=9.8cm, yshift=0.5cm]
            {\textit{4. Evolution of chemical composition}
            \nodepart[align=left]{second} Chemical composition evolution with/without atomic diffusion, nuclear reactions. Adapt mesh used for solving chemical transport equations.};

        \node (converge_2) [rectangle, draw=black, anchor=north, yshift=0.5cm, align=center, text width=3cm, minimum size=8mm, below=of chim] {\small \textit{Convergence ?}};

        \node (QS_1D) [split_cell, below=of converge_2, text width=9.8cm, yshift=0.5cm]
            {\textit{5. Compute new quasi-static structure}
            \nodepart[align=left]{second} Find solution of structure equations. Adapt mesh used for solving structure equations.};

        \node (converge_3) [rectangle, draw=black, anchor=north, yshift=0.5cm, align=center, text width=3cm, minimum size=8mm, below=of QS_1D] {\small \textit{Convergence ?}};

        \node (output) [align=center, yshift=0.5cm, below=of converge_3] {\textit{Next time step}};

        \draw[myarrow] (input.south)      --                    (init.north);
        \draw[myarrow] (init.south)       --                    (limzc.north);
        \draw[myarrow] (limzc.south)      --                    (rot.north);
        \draw[myarrow] (rot.south)        --                    (converge_1.north);
        \draw[myarrow] (converge_1.east)  -- node[above]{no} ++ (4.5,0) node[right]{\parbox{2.5cm}{\centering If no convergence was possible, $\Delta t$ is reduced to $\Delta t = \f{\Delta t}{2}$}} |- (input.east);
        \draw[myarrow] (converge_1.south) -- node[right]{yes}   (chim.north);

        \draw[myarrow] (chim.south)       --                    (converge_2.north);
        \draw[myarrow] (converge_2.south) -- node[right]{yes}   (QS_1D.north);
        \draw[myarrow] (converge_2.east)  -- node[above]{no} ++ (4.5,0) |- (input.east);

        \draw[myarrow] (QS_1D.south)      --                    (converge_3.north);
        \draw[myarrow] (converge_3.south) -- node[right]{yes}   (output.north);
        \draw[myarrow] (converge_3.east)  -- node[above]{no} ++ (4.5,0) |- (input.east);

    \end{tikzpicture}
    \caption{Schematic representation of steps followed by \cesamxx in computing a time step.}
    \label{fig:cesam_flow}
    \end{center}
    \end{figure*}

    The general path followed by \cesamxx to compute the evolution during one time step is summarised in Fig. \ref{fig:cesam_flow}. At a given time step $t + \Delta t$, \cesamxx starts by evaluating the mass lost or gained according to the chosen prescription (see Sect. \ref{sect:mass_loss} for details). Then, based on the structure determined at the previous time step, the new location of the radiative-convective zone (RZ-CZ) transitions is determined, following  Schwarzschild's or Ledoux's criterion, using the method of  \citet{Gabriel2014}  (see. Sect. \ref{sect:conv} for more details). When we know the precise localisation of RZ and CZ, \cesamxx can solve (when needed) equations of the transport of angular momentum (see Sect. \ref{sect:amt} for details). If this step is not a success, the time step is divided by 2 and the process is restarted from the beginning. Otherwise, equations for the transport of chemicals are solved. This includes chemical elements created or depleted by nuclear reactions and diffused by microscopic, turbulent, and radiative diffusion (see Sect. \ref{sect:chim_evol} for details). Again, depending on the convergence of these operations, the process continues or it is restarted with $\Delta t / 2$. Finally, the structure equations are solved using the collocation method described above. Afterwards, in addition to some convergence criteria, we also checked that other criteria were met, depending on the precision needed.

    The numerical precision can be finely adjusted in \cesamxx. In addition to parameters that control B-spline orders, number of layers, repartition factors (see Eq. \eqref{eq:Q_general}), and other numerical aspects, the user can also choose to limit between each time step, the variation of core temperature or density, the variation of $\teff$, $\logg$, and other parameters. This approach can be used to obtain very detailed stellar evolutionary tracks. All these parameters have a default value and a pre-defined set of values can be selected to achieve super precision, solar accuracy, etc. There exists also pre-defined precision for the \corot mission, and for the \plato grid, as \cesamxx was selected to compute the first generation of \plato's grid of stellar models.

    \subsection{Python interface}

    To facilitate the use of \cesamxx, we developed  a Python utility package over the years. This package is divided into two main parts. The first, \texttt{pycesam}, is designed for the running and the post-processing of \cesamxx models. It also creates the interface between two oscillation codes: Adiabatic Pulsation package (ADIPLS; \citep{Christensen-Dalsgaard2008} and the Adiabatic Code of Oscillation including Rotation (ACOR; \citealt{Ouazzani2012}). The second part, \texttt{pycesam.gui} hosts a graphical user interface (GUI). This GUI allows the user to visualize all available options and modify them as they want before running the model. It can also control options of the above-mentioned oscillation code, to compute synthetic oscillation spectrum. This GUI is very adapted for low-intensity model computation work or for teaching. \texttt{pycesam.gui} also provides plot methods used to automatically plot HR diagram, Kippenhahn diagrams, and the profile of various quantities. In addition to these two modules, \cesamxx is also distributed with Python scripts used to build and compute grids of models, compute frequencies in a grid of models, and produce input files for grid-based optimization programs AIMS \citep{Rendle2019} and SPInS \citep{Lebreton2020}.

    \section{Opacity tables and equation of state}
    \label{sect:eos_opac}

    \subsection{Opacity tables}

    \cesamxx implements numerical opacity tables issued by five different collaborations. The first ones are build by the OPAL team \citep{Rogers1992,Iglesias1996}. They exist in two types: Type 1 opacity tables with fixed $Z < 0.1$; and Type 2 opacity tables, which also account for C and O enhancement. The latter are interpolated in \cesamxx with program \texttt{z14xcotrin21.f} written by A. I. Boothroyd\footnote{See \url{https://www.cita.utoronto.ca/~boothroy/}.}. The second ones come from the Opacity Project \citep[OP;][]{Seaton1994,Badnell2005,Seaton2005,Seaton2007}, which offer access to mean opacities, monochromatic opacities and radiative accelerations. The OPAL and OP tables are supplemented at low temperatures by the Wichita\footnote{\url{https://www.wichita.edu/academics/fairmount_college_of_liberal_arts_and_sciences/physics/Research/opacity.php}} opacity tables \citep{Ferguson2005}. They are also adapted to some of the main historical solar abundances determination. Table \ref{table:opa_tables} summarises what is currently available. Above $T = 7\cdot 10^7$ K, the plasma is considered fully ionised, and we only accounted for the Compton scattering by free electrons, evaluated using \cite{Poutanen2017} fitting function, with a smooth transition between $7\cdot 10^7$ K and $8.7\cdot 10^7$ K.

    \begin{table}[t]
        \caption{Opacity tables available in \cesamxx.}
        \label{table:opa_tables}
        \begin{center}
        \begin{tabular}{l L{0.12\textwidth} L{0.11\textwidth}}
            \hline\hline                                                                         \\[-7pt]
            Suffix                                        & Solar abundances & High temperature  \\
            \hline                                                                               \\[-7pt]
            \texttt{GN93}                                 & GN93 (1)         & OPAL              \\
            \texttt{GS98\char`_met}                       & GS98 (2)         & OPAL              \\
            \texttt{GS98\char`_phot}                      & GS98 (2)         & OPAL              \\
            \texttt{AGS09\char`_met}                      & AGS09 (3,4)        & OPAL              \\
            \texttt{AGS09\char`_phot}                     & AGS09 (3)        & OPAL              \\
            \texttt{AGS09\char`_met\char`_OP\char`_WICH}  & AGS09 (3,4)        & OP                \\
            \texttt{AGS09\char`_phot\char`_OP\char`_WICH} & AGS09 (3)        & OP                \\
            \texttt{AAG21\char`_phot}                     & AAG21 (5)        & OPAL              \\
            \hline
        \end{tabular}
        \end{center}
        Notes: Opacity tables based on OPAL or OP tables, available in \cesamxx. The name of each table is \texttt{opa\char`_yveline\char`_} and a suffix. \texttt{met} stands for "meteoritic abundances" and \texttt{phot} for "photospheric abundances". They are all matched linearly in the range $\log_{10} T = 4 \pm 0.05$ K, with Wichita opacity tables at low temperature. (1): \citet{Grevesse1993}; (2): \citet{Grevesse1998}; (3): \citet{Asplund2009}; (4): \citet{Serenelli2009} recommends taking photospheric abundances of \citet{Asplund2009} for valotile elements; (5): \citet{Asplund2021}.
    \end{table}

    In addition to the opacity tables, the user can also choose how \cesamxx will interpolate in these tables. It is advised to use an interpolation scheme validated through the ESTA/CoRoT collaboration \citep[see, e.g.][]{Lebreton2008b,Montalban2008}. Recently, a new interpolation scheme, adapted from the \texttt{xztrin21.f} routine for Type 1 OPAL opacities\footnote{\url{https://opalopacity.llnl.gov/existing.html}} has been added to \cesamxx. It allows for smooth derivatives of the opacity, $\kappa$, to be obtained with respect to the temperature, $T$, or density, $\rho$, suitable for non-adiabatic oscillation computation. Interpolation routines using birational splines from G. Houdek OPALINT\footnote{\url{https://phys.au.dk/~hg62/opint.html}} package are also provided \citep{Houdek1996}.

    \subsection{Conductive opacities}

    Conductive opacities can be computed in three different ways. Analytically, from expressions detailed in \cite{Iben1975}, or numerical by interpolating in tables, either from \citet{Cassisi2007,Potekhin2015} or the updated tables\footnote{\url{www.ioffe.rssi.ru/astro/conduct}} from \citet{Cassisi2021}.

    \subsection{Equations of state}

    Several equations of state (\eos) are available with \cesamxx (see Table \ref{table:eos}). Analytical \eoss are the same as the one presented in ML08. On the side of numerical \eoss, MHD tables \citep{Hummer1988,Mihalas1988} can still be used. Modern \eoss are also provided. The version of the OPAL 2005 \eos \citep{Rogers2002} has been implemented. Recently, a collaborative effort has yielded to the integration of the SAHA-S \eos readily into \cesamxx \citep{Baturin2017}. The SAHA-S \eos \citep{Gryaznov2006,Gryaznov2013} is based on the chemical picture \citep{Ebeling1969} and solar models computed, which seems to be in better agreement with helioseismic data than the one computed with OPAL 2005 (which OPAL 2005 follows the physical picture; \citealt{Vorontsov2013}). SAHA-S is, at the moment, only adapted for $0.1 \leq X \leq 0.9$, which prevents its use for stellar models beyond the main sequence or in specific cases, such as chemically peculiar stars with high hydrogen content at the surface. Routines adapted to SAHA-S are readily available in \cesamxx and the user should only download directly the data from the SAHA-S website\footnote{\url{http://crydee.sai.msu.ru/SAHA-S_EOS/index.php}}.

    \begin{table}[t]
        \caption{Summary of the EoSs available in \cesamxx.}
        \label{table:eos}
        \begin{center}
        \begin{tabular}{l l l l}
            \hline\hline                                                            \\[-7pt]
            \eos      & Kind       & Picture  & \texttt{nom}\char`_\texttt{etat}    \\
            \hline                                                                  \\[-7pt]
            GONG1 (1) & Analytical & Chemical & \texttt{etat}\char`_\texttt{gong1}  \\
            GONG2 (2) & Analytical & Chemical & \texttt{etat}\char`_\texttt{gong2}  \\
            EFF (3)   & Analytical & Chemical & \texttt{etat}\char`_\texttt{eff}    \\
            CEFF (4)  & Analytical & Chemical & \texttt{etat}\char`_\texttt{ceff}   \\
            MHD       & Numerical  & Chemical & \texttt{etat}\char`_\texttt{mhd}    \\
            OPAL 2005 & Numerical  & Physical & \texttt{etat}\char`_\texttt{opal5Z} \\
            SAHA-S    & Numerical  & Chemical & \texttt{etat}\char`_\texttt{saha}   \\
            \hline
        \end{tabular}
        \end{center}
        Notes: Last column corresponds to the value of the option \texttt{nom}\char`_\texttt{etat} in an input file. (1) \citet{Christensen-Dalsgaard1988} with only H and He, assuming complete ionization, and neither the degeneracy nor ionization internal energy or radiation pressure; (2) \citet{Christensen-Dalsgaard1988} with only H and He, assuming complete ionization, and neither the degeneracy nor radiation pressure; (3) \citet{Eggleton1973}; (4) \citet{Christensen-Dalsgaard1992}.
    \end{table}

    \section{Convection}
    \label{sect:conv}

    The modelling of convection is one of the major locks for our understanding of stellar structure and evolution. We thus carried many developments to improve the accuracy of the models. The two most important changes are  1) a new scheme for the determination of mixing region boundaries and (2) the possibility to have the convection parameter varying along evolution and physically constrained through the entropy calibration.

    \subsection{Determination of limits with \citet{Gabriel2014}}

    \begin{figure*}[t]
        \begin{center}
            \includegraphics[draft=False, width=\textwidth]{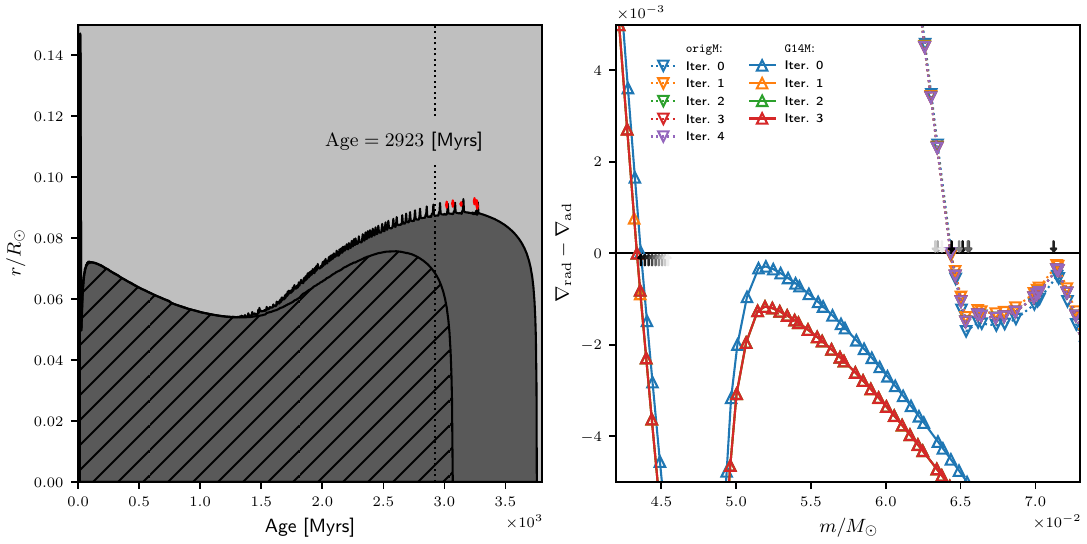}
            \caption{Left panel: Kippenhahn diagram of two models of $1.3M_\odot$ computed with the method originally implemented in \cesam \citep[hereafter model \texttt{OrigM}]{Morel1997}, and with the method prescribed by \citet[hereafter model \texttt{G14M}]{Gabriel2014} and now implemented in \cesamxx. The radiative zone corresponds to dark grey region for \texttt{OrigM}, and hatched region for \texttt{G14M}. Convective zone is the region in light grey for \texttt{OrigM}, and non-hatched region for \texttt{G14M}. Red regions are spurious convective zone. Vertical dashed line at age $2923$ Myrs marks the position in time of the model represented on right. Right panel: Profile of  $\dd\nabla \equiv \nabla_{\rm rad} - \nabla_{\rm ad}$ as a function of the mass coordinate, $m/M_\odot$, at each iteration of a given time step (at age $2923$ Myrs). Profiles for \texttt{G14M} are drawn in solid lines, and profiles for \texttt{OrigM} are in dashed lines. Up (\texttt{G14M}) or down (\texttt{OrigM}) triangles mark the location of each layer, at each iteration. Grey up (\texttt{G14M}) or down (\texttt{OrigM}) arrows mark the location of the convective boundary at the ten previous time steps. Lightest arrow if for the tenth previous time step and darkest arrow for the immediate previous one.}
            \label{fig:lim_zc_orig_g14}
        \end{center}
    \end{figure*}

    One of the major improvements in this version of \cesamxx is the new method used to determine the convective boundaries. We define
    \begin{equation}
        \dd\nabla \equiv \nabla_{\rm rad} - \nabla_{\rm ad} ~ \stp{+ \frac{\varphi}{\delta} \nabla_\mu},
    \end{equation}
    the Schwarzschild's criterion (Ledoux criterion if the term between parenthesis is added), where $\nabla_{\rm rad}$, $\nabla_{\rm ad}$, and $\nabla_\mu$ are the radiative, adiabatic, and mean molecular weight gradients. Originally, \cestam was computing the $\dd\nabla$ profile and between the two consecutive layers, where $\dd\nabla$ changes sign, the boundary is found by a linear interpolation. This process was repeated at each iteration during a given time step (see step 2 in Fig. \ref{fig:cesam_flow}). This technique was known to produce very erratic results, especially for models with a recessing convective core, where a mean molecular weight gradient has built up. In this case, $\dd\nabla$ becomes discontinuous at the boundary.

    Figure \ref{fig:lim_zc_orig_g14} displays on the left a Kippenhahn diagram on which the dark grey region represents, for a $1.3M_\odot$ model called \texttt{OrigM}, the extent of the core CZ through time, as determined by the original method, and in light grey the RZ above it. From $\sim 1.5~\mbox{Gyrs}$ to the vanishing of the core CZ, the location of the limit appears to be oscillating more and more. After $\sim 3.0~\mbox{Gyrs}$, we can even distinguish the temporary apparition of spurious CZ in red, just above the core CZ. These zones are only consequences of the numerical scheme, but have no physical meaning. The right panel shows the profiles of $\dd \nabla$, in dashed lines for \texttt{OrigM}, around the location of the limit CZ/RZ, for successive iterations performed during the computation of a time step. Grey arrows represent the location of the limit RZ/CZ at the ten previous time steps: in light grey, the location at the tenth previous time step, in dark grey at the immediate previous one. The method originally implemented in \cesamxx predicts a limit that does not decrease monotonically, for a convective core normally recessing at this stage of evolution

    According to \citet{Gabriel2014} who examined in detail the possible misuses of boundary conditions, the location of the boundaries should not be found by looking for a change of sign $\dd \nabla$ between two consecutive layers by scanning from centre to surface. In case of a discontinuity, in the chemical composition, a change of sign can occur while the two layers are both radiative, both convective or radiative on one side, convective on the other. The only way to properly locate the boundary at $\dd\nabla = 0$ is to always extrapolate from the interior of a convective zone. \citet{Gabriel2014}  also advocates for the use of a double mesh point (two points located at the same mass coordinate: one in RZ and the other one in CZ), in case of a discontinuity. This was already implemented in \cesam and \cestam as adding extra mesh points at discontinuities is part of the B-Spline machinery.

    The limit of the core CZ/RZ for a $1.3M_\odot$ model called \texttt{G14M} obtained with the \citet{Gabriel2014} method is displayed in Fig. \ref{fig:lim_zc_orig_g14} (left panel). The extent of core CZ through time corresponds to the hatched region. The limit evolved in a much smoother way compared to what was obtained for \texttt{OrigM}. The right panel shows the profiles of $\dd \nabla$ for \texttt{G14M} in bold line  through successive iterations at a given time step. The grey arrows show, as expected, that the boundary of the recessing core CZ is smoothly moving inwards through time. This new scheme is also more stable and converges faster than the original one. For the considered time step, convergence is reached after four iterations, while five were  needed in the past. The new method is not without consequences for evolution. Because of the very erratic limit locations found with the old scheme, more hydrogen was injected from RZ to core CZ, thus prolonging the stellar lifetime. As an example, model \texttt{OrigM} reaches the TAMS (defined as central $X$ reaching $0.01$) at an age of $3740$ Myrs, while model \texttt{G14M} reaches it at $3060$ Myrs, a difference of around $20\%$.

    \subsection{Entropy calibration}

    The issues with ad hoc models of convection used in 1D stellar evolution code have been known for some time and well studied, even if (at the moment) we lack better alternatives. For a detailed review of these problems, see \citet{Trampedach2010} and for a review on convection modelling, see \citet{Kupka2017}. At the moment, the most widely used prescriptions for modelling convection are  mixing length theory \citep[MLT;][]{Biermann1932,Siedentopf1935,Biermann1942,Bohm-Vitense1958}, where the turbulence spectrum is reduced to a Dirac distribution (i.e. all convective flux is carried by one single eddy), and full spectrum of turbulence models \citep[FST;][]{Canuto1991,Canuto1992,Canuto1996} that approximate the turbulence spectrum with a Kolmogorov spectrum. As far as we know, \cesamxx is the only stellar evolution code with ATON Rome Stellar Evolution Code \citep{Ventura1998} that implement all these models \citep[see,][]{Joyce2023}. MLT and FST all rely on a free parameter called the convection parameter, $\alpha$. Its definition slightly differs from one theory to the other, but its role is always to tune the properties of the convective flow so that the stellar model reproduces some observables. The choice of $\alpha$ has always been a challenge as it is not physically constrained. Yet its value strongly impact the evolutionary track and, in turn, stellar fundamental parameters estimates. Except in the case of a solar model where observable constraints are sufficiently tight to often ensure that a single value of $\alpha$ allows reproducing them, in case of other stars the solution is often degenerate. In addition, $\alpha$ is always assumed to be constant through time, while there is no good reason to think that properties of the convective flow should not vary across evolution.

    To circumvent these issues, two approaches have been developed. The first one is to compute 1D stellar atmosphere models by tuning, among other parameters, $\alpha,$ to reproduce global properties of the 3D, more realistic, models of surface convection; for instance, the effective temperature $\teff$, $\logg$ and $\feh$. Once a sufficient number of such 1D models have been obtained, it allows calibrating a function $f(\teff,\logg,\feh)$ that fits the $\alpha$ values in the grid. Such a function can then readily be used in 1D stellar evolution codes to set the value of $\alpha$ for any tuple $(\teff,\logg,\feh)$. This approach has been followed by \citet{Sonoi2019} who provide six fitting functions for six different set of microphysics, all implemented in Cesam2k20. With such a method, $\alpha$ is not fixed along evolution and its value is not a free parameter any more. However, a disadvantage is that the microphysics chosen to compute one's model should match exactly the one used to compute the fitting function, which advocates for the second approach.

    The second approach uses prescriptions for the value of the specific entropy of the adiabat of the convective envelope. Since the value of $\alpha$ directly determines on which adiabat the star is on, \citet{Spada2018,Spada2019,Spada2021} suggested to simply tune $\alpha$ along the evolution such that, at any time, the specific entropy of the adiabat of the 1D matches a prescribed value. This idea was implemented in \cesamxx and further tested by \citet{Manchon2024}, who also explain how to correct the entropy prescriptions for different chemical compositions or choice of EoS, and they give recommendations on the use of prescriptions already provided by \citet{Ludwig1999,Magic2013,Tanner2016}. This approach has several advantages over the first one. First, it relies on prescriptions for the specific entropy, which has a physical meaning, while $\alpha$ depends on the microphysics of the model.  Second, it suppresses the need of choosing a poorly defined parameter when fitting a star, and allows either to remove a degree of freedom, or to calibrate an additional parameter that controls another physical process. Finally, EC modelling authorizes $\alpha$ to vary along evolution, which seems a strong approximation of the classical MLT modelling.

    \subsection{Overshoot}

    The effect of the overshoot is taken into account since the earliest version of \cesam. The overshooting regions were first assumed to be fully mixed and recently, the possibility to assume a diffusive mixing was added, according to the formulation of \citet{Herwig2000}, with the velocity scale height taken as $H_{\rm v} = \alpha_{\rm ov}H_{\rm p}$, where $\alpha_{\rm ov}$ is the overshoot parameter and $H_{\rm p}$ the pressure scale height. In \cesamxx, the overshoot parameters of the convective core ($\alpha_{\rm ov,c}$) and of the envelope ($\alpha_{\rm ov,e}$) can be controlled independently. The size of the overshooting region over the convective core is then ${\rm min}(\alpha_{\rm ov,c}H_{\rm p}; R_{\rm c} \alpha_{\rm ov,c})$, with $R_{\rm c}$ the radius of the convective core, and is $\alpha_{\rm ov,e}H_{\rm p}$ below the convective envelope.

    Several options can be chosen for setting the temperature gradient $\nabla$ in the overshooting region. By default, we assume an adiabatic temperature gradient above the convective core, and a radiative gradient (or adiabatic if recommendation of \citealt{Zahn1991} are followed) below the convective envelope. \citet{Rempel2004} show that these simple prescriptions lead to shallow overshooting regions, with sharp transitions in the temperature gradient, in contradiction with numerical simulation.  To overcome this issue, \citet{Christensen-Dalsgaard2011b} proposed an analytical expression for the temperature gradient in the overshoot region as
    \begin{equation}
        \nabla = \nabla_{\rm ad} - \frac{2\mF_{\rm ov}}{\beta + \exp( 2\zeta )},
    \end{equation}
    where $\mF_{\rm ov}$, $\beta$, and $\zeta$ are certain quantities that control the shape of the transition (see \citealt{Christensen-Dalsgaard2011b} for details). This formulation allows to better reproduce the size of the penetrative convection region as inferred by helioseismology. However, for hotter stars, in which the radiative gradient does not decrease monotonically below the Schwarzschild's boundary, a more complex form for $\nabla$ can be considered, such as the one introduced by \citet{Deal2023}:
    \begin{equation}
        \nabla = \nabla_{\rm ad} - \frac{\nabla_{\rm ad} - \nabla_{\rm rad}}{2}\left[1 - \frac{2}{\pi} \arctan \left(\frac{\zeta(r) - \alpha_{\rm PC} H_p(r_{\rm e})}{\beta(\nabla_{\rm ad} - \nabla_{\rm rad})^4}\right)\right],
        \label{eq:nabla_morgan}
    \end{equation}
    with $\zeta(r) = r - r_{\rm e}$, $r_{\rm e}$ being the radius of the Schwarzschild's boundary of the convective envelope, $\alpha_{\rm PC}$ is a tunable parameter, and $\beta$ controls the steepness of the transition. All these options are available in \cesamxx for setting the gradient in the overshooting region.

    \section{Evolution of the chemical composition}
    \label{sect:chim_evol}

    From the first version of \cesam to the current \cesamxx, a great deal of attention has been paid on the evolution of the chemical composition, and especially, the transport of chemical elements. This has contributed, over the years, to the acknowledgement of the chemical diffusion as being part of the standard model of stellar physics. The evolution of the concentration of an element $X_i$ is governed by nuclear reactions and mixing processes such as convective motions, gravitational settling, radiative acceleration or MHD instability-induced mixing. This can be expressed by the following equation,
    \begin{eqnarray}
        \rho \pd{X_i}{t} &=&-\lambda_{i,\rm nuc} \rho X_i - \frac{1}{r^2} \pd{}{r}\stp{r^2\rho v_{i,\rm H} X_i} \nonumber \\
                         & &- \frac{1}{r^2} \pd{}{r}\stp{r^2\rho D_{\rm turb} \pd{X_i}{r}},
        \label{eq:tce}
    \end{eqnarray}
    where $\lambda_{i,\rm nuc}$ is the nuclear reaction rate of specie $i$, $v_{i,\rm H}$ is the atomic diffusion velocity of specie $i$ relative to hydrogen, and $D_{\rm turb}$ is a coefficient of diffusion associated with the turbulence. This general system of equations is solved using a relaxation technique \citep{Press1995}. Of course, if the user chooses to neglect atomic and turbulent diffusion, only the first term of the right-hand side remains.

    \subsection{New nuclear reaction rates}

    The nuclear reaction rates presented in ML08 are still available. In addition, the rates obtained from the compilations of NACRE II \citep{Xu2013} -- possibly supplemented with the rate of LUNA \citep{Broggini2018} for the ${}^{14}{\rm N(p},\gamma){}^{15}{\rm O}$ reaction -- or from the Reaclib database \citep{Cyburt2010} can be chosen. With \cesamxx, the user can only follow the evolution of elements that are part of a predefine nuclear networks, among which the user can choose the one that best suits their needs. The abundances of elements left out of the network are assumed to stay at equilibrium during evolution. These networks can account for reactions of the PP chains, CNO cycle or triple-$\alpha$ process with rates chosen from the above compilations. Screening effect can be included according to \citep{Clayton1968}'s prescription for weak screening (default) or following \citet{Mitler1977}'s formalism. As mentioned in the introduction of this section, when atomic diffusion is neglected, Eq. \eqref{eq:tce} simplifies. In this case, the system is solved using an implicit Runge-Kutta type Lobatto IIIC integration scheme for stiff problems (see \citealt{Hairer1996} or ML08 for more details).

    \subsection{Atomic diffusion}

    To compute atomic diffusion velocities, two formalisms are available. The first one consists in solving the full Burger's equations for each element \citep{Burgers1969}, and the second one, following \citet{Michaud1993} amount to consider all elements other than hydrogen as test particles that only diffuse with respect to protons.

    To the atomic diffusion velocities, one can add a contribution coming from the radiative accelerations. \cesamxx allows the use of the Single-Valued Parameter method (SVP; \citealt{Alecian2002,Alecian2004}) or its update from \citet{Alecian2020}. SVP method allows, in an optically thick medium, to separate the properties of the atoms and of the plasma. With such simplification, and assuming that the probabilities of bound-bound transitions (resp. bound-free transitions) have a Lorentzian profile (resp. Voigt profile), the expression for the radiative acceleration of a given element depends only on six parameters specific to the element, and its local abundance. This dramatically speeds up the computation with a small associated error.

    \subsection{Ad hoc turbulent diffusion prescriptions}

    The last term of Eq. \eqref{eq:tce} gathers all the contributions to the atomic diffusion induced by the turbulence. This includes known processes such as double diffusive convection (see Sect. \ref{subsection:ddc}) and rotation-induced mixing (see Sect. \ref{sect:amt}). Other options exist within \cesamxx to parametrise turbulent mixing with ad hoc prescriptions. Two of them, already available in later version of \cesam rely either on a coefficient calibrated below the Solar CZ \citep{Gabriel1997} or a fixed value used in regions where $T < 10^6$ K, avoiding Helium's sedimentation.

    More recently, we have implemented other prescriptions. A first type of these have a turbulent diffusion coefficient of the form,
    \begin{equation}
        D_{\rm turb} = \omega D({\rm He})_0 \stp{\frac{\rho_0}{\rho}}^n,
        \label{eq:dturb_montreal}
    \end{equation}
    where $\omega$ and $n$ are two constants which can be set in the input file of \cesamxx, $D({\rm He}),$ and $\rho$ are the local diffusion coefficient of Helium and density, while the subscript $0$ indicates that the value is taken at a reference depth. This reference depth can be set at a given mass fraction or temperature (prescription of \citealt{Richer2000}) or at the base of the convective zone (see the prescription of \citealt{Proffitt1991}). The reference depth can also be varied between two calibrated reference values that allow us, for example, to reproduce the helium abundance of F-type stars \citep{Verma2019} and the ${\rm Li}^7$ of the Sun, following an idea applied in \citet{Moedas2025}. The second type of prescription \citep{Freytag1996,Theado2009,Deal2016,Deal2018} takes the following form,
    \begin{equation}
        D_{\rm turb} = \omega_1 \exp\stp{ \frac{r - r_{\rm e}}{\sigma_1 R}\ln 2 } + \omega_2 \exp\stp{ \frac{r - r_{\rm e}}{\sigma_2 R}\ln 2 },
    \end{equation}
    where $\omega_1$, $\omega_2$, $\sigma_1$ and $\sigma_2$ are user-defined constants, and $r_{\rm e}$ was already defined in Eq. \eqref{eq:nabla_morgan}.

    \subsection{Double diffusive convection}
    \label{subsection:ddc}

    To this ad hoc treatment of turbulent diffusion, \cesamxx is also able to include the effect of specific physical mechanisms, namely, the thermohaline convection or the semi-convection. The first one is triggered in the region where the medium is stable with respect to the thermal gradient,  but it is unstable because of the gradient of mean molecular weight. Thus, it is treated using the formalism ofrom \citet{Brown2013}. The other is triggered when opposite conditions are met and \cesamxx models it following \citet{Langer1983}.

    \section{Evolution of the angular momentum distribution}
    \label{sect:amt}

    \subsection{Angular momentum transport according to \citet{Zahn1992}}

    The treatment of angular momentum transport according to \citet{Zahn1992} in \cesamxx is described in M13. We recall the general ideas and then report on the new, non-standard developments. Angular momentum is globally conserved in a \cesamxx model except that it can be gained at the surface through planetoid infall (see ML08), or lost at the surface through magnetized stellar winds. While the analytical model of \citet{Kawaler1988} is still available, \cesamxx now offer the possibility to use a scaling relation derived from 3D simulation of magnetized stellar winds \citep{Matt2015}, which assumes a more complex geometry of the magnetic field.

    In the interior, the treatment of the transport of angular momentum (TAM) is different whether we are in a convective or radiative zone. Inside the former, because of the efficient mixing, we can assume uniform distribution of the angular velocity (solid body rotation), or of the angular momentum (local conservation of angular momentum). Inside a radiative zone, we use the framework of \citet{Zahn1992} and \citet{Talon1997b}. First, large scale meridional currents are needed to verify baroclinic equilibrium. This meridional circulation advects angular momentum $\mathcal{J}$ and chemical elements. Second, shear-induced turbulence (Kelvin-Helmoltz instability) diffuses angular velocity $\Omega$. Third, since this viscosity is much stronger horizontally than vertically, we assume that the radiative zone is in shellular rotation, namely, $\Omega$ depends only on the radial coordinate. The TAM is then modelled in \cesamxx as an advecto-diffusive process. The latitudinal variations are decomposed as a series of Legendre polynomials, and degrees $\ell > 2$ are neglected. Hence, the angular momentum transported vertically follows
    \begin{equation}
        \rho\diff{r^2\Omega}{t} = \frac{1}{5r^2}\pd{}r{} \stp{\rho r^4 \Omega U_2} + \frac{1}{r^2} \pd{}{r}\stp{\rho \nu_{\rm v} r^4 \pd{\Omega}{r}},
        \label{eq:tam}
    \end{equation}
    where $U_2$ is the vertical component of the meridional circulation, and $\nu_{\rm v}$ is the vertical shear-induced viscosity coefficient. Equation \eqref{eq:tam} is solved with the linear two-point boundary relaxation method described in \citet{Press1995}. Coefficients of the differential equations are computed explicitly with the solution of the structure found at previous iteration. Many prescriptions have been suggested to provide a value for the diffusion coefficients of the Kelvin-Helmholtz instability, either from theoretical considerations or experiments. \cesamxx implements three of them as proposed in \citet{Palacios2003,Mathis2004b} and \citet{Mathis2018}.

    The effect of the TAM on the transport of chemicals is accounted for through an additional term of the right-hand side of Eq. \eqref{eq:tce}, expressed as
    \begin{equation}
        \frac{1}{r^2}\pd{}{r}\stb{r^2 \rho (D_{\rm v}+D_{\rm eff}) \pd{X_i}{r}}
    ,\end{equation}
    where $D_{\rm v} = \nu_{\rm v}$ and $D_{\rm eff} = (rU_2)^2 / (30D_{\rm h})$, with $D_{\rm h}$ the horizontal shear-induced viscosity coefficient.

    Finally, to initialize the distribution of angular momentum, we make use of the disc-locking model. As suggested by \citet{Bouvier1997},  young, completely convective, fast-rotating stars have  strong magnetic fields that lock the convective zone with the accretion disc. Then, the convective envelope of the star is forced to co-rotate with the accretion disc, as long as it exists. With only two parameters being the rotation period, $P_{\rm disc}$, and its lifetime, $\tau_{\rm disc}$, this model provides a simple initial distribution of angular momentum.

    \subsection{Additional transport mechanisms}

    It has been known for more than a decade that this model fails to reproduce the flat internal rotation profile of the solar radiative zone derived from helioseismology. The core rotation of subgiant and red giant stars derived from asteroseismic observations also strongly disagrees with the model's predictions \citep[e.g.][]{Eggenberger2012,Marques2013,Goupil2013b,Cantiello2014,Fuller2014,Ouazzani2019}. Several mechanisms have been suggested to explain the missing TAM. These mechanisms fall into two categories: either angular momentum advected by waves, or a diffusion of angular velocity by (magneto)hydrodynamic instabilities. At the moment, none of them is found to solve the TAM problem alone.

    \cesamxx can account for the angular momentum transported by mixed-modes, following the idea developed by \citet{Belkacem2015a,Belkacem2015b}. This is done by adding a flux term $-1/r^2 \lpd{(r^2 \mathcal{F}_{\rm waves})}{r}$ in the right-hand side of \eqref{eq:tam}. While \citet{Belkacem2015b} computed the fluxes a posteriori for a few models, with eigenfunctions computed using ADIPLS \citep{Christensen-Dalsgaard2008}, transport by mixed-modes can now be computed throughout the evolution, using asymptotic expression for eigenfrequencies and eigenfunctions, as well as a scaling relation for the amplitude of the eigenmodes \citep[e.g.][]{Mosser2010,Samadi2011,Belkacem2013a,Benomar2014}.

    On the side of (magneto)hydrodynamic instabilities, two of them have been recently added or improved in \cesamxx. First, the Goldreich-Schubert-Fricke (GSF) instability \citep{Goldreich1967,Fricke1968,Acheson1978}, which is purely hydrodynamical, occurs when the shear perpendicular to the rotation axis is strong enough to overcome the stabilizing effect of the thermal and chemical stratifications. In addition to the model that already existed in \cestam (see M13), we now also propose to use the GSF coefficient of angular momentum diffusion obtained from scaling relations extracted from results of 2D simulations of GSF instability near the stellar equator presented in \citet{Barker2019}.

    The Tayler instability (TI) is thought, at the moment, to be among the most efficient MHD instability at transporting angular momentum. The latest version of \cestam already included the TI, following the analytical formalism of \citet{Spruit2002} generalised by \citet{Maeder2004}. Due to very short timescale of diffusion of angular velocity compared to the timescale of the evolution of the structure, TI is known to induce numerical errors \citep[see e.g.][]{Yoon2004,Neunteufel2017,Wheeler2015}. Recently, new analytical model of TI as proposed by \citet{Fuller2019} or scaling relations derived by \citet{Daniel2023} from 3D MHD simulations have been added to \cesamxx. Moreover, a lot of effort have been put into improving its implementation and this will be the topic of a forthcoming paper. With the current version of \cesamxx, models including the TI as a transport process will produce angular velocity profiles that might be reliable qualitatively but are still impacted by numerical artefacts.

    \section{Atmosphere}
    \label{sect:atmosphere}

    The details of the implementation of atmosphere reconstruction in \cesamxx have already been described in ML08, and even more precisely in \citet{Morel1994}. Two possibilities exists to model the atmosphere:  using a $\ttau$ relation or interpolation into a pre-computed grid of stellar atmospheres.

    \subsection{New purely-radiative $\ttau$ relations}

    In the atmosphere, the system of structure equations can be re-written with the optical depth, $\tau$, as the reference coordinate. To close the system, the temperature profile can be set through a $\ttau$ relation,
    \begin{equation}
        \ttau = \teff\stb{\frac{3}{4}\stp{q(\tau) + \tau}}^{1/4},
    \end{equation}
    where $q(\tau)$ is the Hopf function, the value of which must be prescribed. Three new $\ttau$ relations have been added to the new version of \cesamxx (see. Table \ref{table:ttau}. The first two, derived by \citet{KrishnaSwamy1966} and \citet{Vernazza1981}, are constructed to reproduce, respectively, observations of K dwarfs, and of the Sun's photosphere (called VAL-C model). The last $\ttau$ relation has a functional form suggested by \citet{Ball2021}, and is calibrated on a 3D simulation of the Sun \citep{Trampedach2014a}. Its behaviour is quite similar to the VAL-C model for large $\tau$, but reproduce better the shape of $q(\tau) \propto \log \tau$ in the limit of $\tau \rightarrow 0$.

    \begin{table}[t]
        \caption{Summary of $\ttau$ relations available in \cesamxx.}
        \label{table:ttau}
        \begin{center}
        \begin{tabular}{l l l}
            \hline\hline                                                        \\[-7pt]
            Name               & \texttt{nom}\char`_\texttt{tdetau} & Comments     \\
            \hline                                                                 \\[-7pt]
            Eddington          & \texttt{edding}  & Purely radiative.  \\
            Hopf               & \texttt{hopf}    & Purely radiative.  \\
            Krishna-Swamy      & \texttt{krisw}   & Purely radiative.  \\
            VAL-C              & \texttt{verna}   & Purely radiative.  \\
            Hypergeometric fit & \texttt{ball21}  & Purely radiative.  \\

            \hline
        \end{tabular}
        \end{center}
        Notes: \texttt{nom}\char`_\texttt{tdetau} corresponds to the name of the variable in the code.
    \end{table}

    \subsection{Non-purely-radiative $\ttau$ relations}

    Non-purely-radiative $\ttau$ relations are the same as the one already mentioned in ML08 paper, only the corresponding tables have been updated. Several routines based on the Kurucz's ATLAS9 non-radiative model atmospheres \citep[see, e.g,][and their  description of the different versions of ATLAS models and references therein]{Kurucz2014} were originally written to model the atmosphere of the Sun \citep[see e.g.][]{Morel1994}, but are no longer used. Then, a more general routine \texttt{rogerYL} has been designed to allow interpolation within a large set of $T(\tau)$ laws derived from ATLAS12 models. These laws cover the range $3500\, \mathrm{K} < T_\mathrm{eff} < 7000\, \mathrm{K}$ and $1 < \logg < 5$ and are provided for a solar mixture and four values of [Fe/H] ($-1.0, -0.5, +0.0, +0.2$, for the value $-1.0$ the law takes into account an $\alpha$-element enrichment of $+0.4$ characteristic of low metallicity stars). In a former version of CESAM, $T(\tau)$ laws from the MARCS \citep{Gustafsson2008} and PHOENIX \citep{Allard2014} model atmospheres have been implemented; they will soon be available in the present version.

    \section{Stellar mass loss}
    \label{sect:mass_loss}

    Very recently, a lot of work has been put into better describing the mass loss in stellar models. \cesamxx implements eight different mass-loss prescriptions and can switch from one to the other along  the evolution. The most simple prescriptions involve a constant mass loss (or gain), $\dot{M}$. More complex ones are derived from observations or 3D simulations. A brief summary of available options is given in Table \ref{table:ml}.

    {\renewcommand\arraystretch{1.75}
    \begin{table*}[t]
        \caption{Summary of mass-loss recipes available in \cesamxx.}
        \label{table:ml}
        \begin{center}
        \begin{tabular}{L{0.15\textwidth} L{0.5\textwidth} L{0.25\textwidth}}
            \hline\hline
            Name                                  & Comments       & Can be used with    \\
            \hline
            \texttt{pert}\char`_\texttt{ext}      & \begin{minipage}[t]{0.5\textwidth}
                Constant mass-loss (or gain).
                                                    \end{minipage} & Always \\
            \texttt{pert}\char`_\texttt{vink}     & \begin{minipage}[t]{0.5\textwidth}
                Mass-loss depending on $L$, $M$, $\teff$, $Z$ and terminal velocity of the wind $v_\infty$ \citep{Vink2001}. A specific formulation for the bi-stability jump is also implemented.
                                                    \end{minipage} & Hot stars ($\teff > 10^4$ K) \\
            \texttt{pert}\char`_\texttt{Reimers}  & \begin{minipage}[t]{0.5\textwidth}
                Mass-loss proportional to $LR/M$ \citep{Reimers1975,Maeder1989}. Proportionality factor is calibrated on observations, but slightly varies with stellar mass and evolution stage.
                                                    \end{minipage} & \begin{minipage}[t]{0.25\textwidth}Solar-type, SG, RGB, AGB, supergiants \end{minipage} \\
            \texttt{pert}\char`_\texttt{vanLoon}  & \begin{minipage}[t]{0.5\textwidth}
                Mass-loss depending on $L$ and $\teff$ for oxygen-rich and pulsating AGB and red supergiant stars \citep{vanLoon2005}.
                                                    \end{minipage} & SG, RGB, AGB, supergiants \\
            \texttt{pert}\char`_\texttt{SC05}     & \begin{minipage}[t]{0.5\textwidth}
                Improved Reimers-like law, physically grounded \citep{Schroeder2005}. Two more quantities are involved in the expression of the mass-loss: $\teff$ and the gravity. Such an expression better reproduces observed mass-loss rates, in particular mass-loss rates of supergiants. The proportionality factor varies much less from one star to the other that in \citet{Reimers1975}'s original formulation.
                                                    \end{minipage} & SG, RGB, AGB, supergiants \\
            \texttt{pert}\char`_\texttt{Rosenf14} & \begin{minipage}[t]{0.5\textwidth}
                \citet{Rosenfield2014} improves the work of \citet{Schroeder2005} with a better agreement with evolved stars (TP-AGB stars which experience a series of He-shell flashes).
                                                    \end{minipage}  &  AGB, supergiants \\
            \hline
        \end{tabular}
        \end{center}
        Notes: They can be set to the options \texttt{nom\char`_pertm}, \texttt{nom\char`_pertm\char`_solar}, \texttt{nom\char`_pertm\char`_rgb}, \texttt{nom\char`_pertm\char`_agb}, \texttt{nom\char`_pertm\char`_hot}, depending on the star type or evolutionary stage.
    \end{table*}}

    Apart from simple recipes, which account for a constant mass loss or gain during part or the entire lifetime of the star, all others implemented in \cesamxx follow a Reimers-like law \cite{Reimers1975}, expressed as    \begin{equation}
        \dot{M} \propto \eta \frac{LR}{M},
    \end{equation}
    where $\eta$ is a proportionality factor that should be the same for different stars and different evolutionary status, $L$ is the luminosity, $R$ the radius, and $M$ the mass. In the original formulation by \citet{Reimers1975}, $\eta$ was actually found to vary with stellar mass and evolutionary stage. Later works slightly changed this original formulation to obtain a constant $\eta$ \citep{Schroeder2005,Rosenfield2014}. In the end, it seems that a better approach is to switch from one formulation to the other as the star gets older, which is now allowed in \cesamxx.

    \section{Application: Solar models}
    \label{sect:nssm}

    To demonstrate the abilities of \cesamxx, we computed a series of standard and non-standard solar models. We did not choose to show examples of all the new options, but, rather, we illustrate the main progresses of the code. These models are computed with the help of the OSM package\footnote{\url{https://pypi.org/project/osm/} provided with \cesamxx. Original developer: Réza Samadi; Version for binary stars; L. Manchon.}, that finds the \cesamxx model that best matches a set of observational constraints, by tuning a set of adjustable parameters, following a Levenberg-Marquardt algorithm. The set of these parameters varies from one model to the other, as well as the numerical precision, and the input physics.

    \subsection{Physical inputs}

    \paragraph{\textit{Common physical ingredients.}} \label{paragraph:comphys} All models were stopped at the same age, $4570~\mbox{Myrs}$ and computed with the same reference composition ratios of \citet[][hereafter AAG21]{Asplund2021}, following the 'meteoritic' compilation, \eos (OPAL 2005), opacities (OPAL adapted to the AAG21), and MLT. The adopted preset numerical precision are named after the space-borne mission for which they were tailored: \texttt{'co'} for CoRoT and \texttt{'pl'} for PLATO. Thanks to more modern computational resources, the PLATO precision is higher than the CoRot one. In particular, it introduces more shells and more time-steps during evolution. We describe below the main characteristics of the solar models. These are summarised in Table \ref{table:snssm_phy}, while the set of observational constraints and the optimal values of tunable parameters are listed in Table \ref{table:snssm_obs} and Table \ref{table:snssm_par}.

    {\renewcommand\arraystretch{1.75}
    \begin{table*}[t]
        \caption{Physical ingredients of the standard and non-standard solar models. }
        \label{table:snssm_phy}
        \begin{center}
\begin{tabular}{L {0.15\textwidth} l L {0.14\textwidth} L {0.14\textwidth} L {0.14\textwidth} L {0.14\textwidth}}
    \hline\hline
    Name                      & SSM1            & SSM2                          & NSSM1                         & NSSM2                         & NSSM3      \\
    \hline
    Adjustable Parameters     & $Y_0$, $\amlt$  & $Y_0$, $(Z/X)_0$,
                                                  $\amlt$, $\alpha_{\rm ov, e}$ & $Y_0$, $(Z/X)_0$, $\amlt$,
                                                                                  $T_{\rm mix}$                 & $Y_0$, $(Z/X)_0$, $\amlt$,
                                                                                                                  $T_0$                        & $Y_0$, $(Z/X)_0$, $\amlt$,
                                                                                                                                                  $T_{\rm mix}$, $T_0$                     \\
    Observational constraints & $\teff$, $L$    & $\teff$, $L$, $(Z/X)_\ss$,
                                                  $T_{\rm cz}$                  & $\teff$, $L$, $(Z/X)_\ss$,
                                                                                  $\log \epsilon_{\rm Li,s}$            & $\teff$, $L$, $(Z/X)_\ss$,
                                                                                                                  $\Omega_\ss$                  & $\teff$, $L$, $(Z/X)_\ss$,
                                                                                                                                                  $\log \epsilon_{\rm Li,s}$,
                                                                                                                                                  $\Omega_\ss$                            \\
    Numerical precision       & \texttt{'co'}   & \texttt{'pl'}                 & \texttt{'pl'}                 & \texttt{'pl'}                 & \texttt{'pl'}                           \\
    Envelope overshoot        & No              & Yes                           & No                            & No                            & No                                      \\
    Gravitational settling    & No              & MP93                          & MP93                          & MP93                          & MP93                                    \\
    Radiative acceleration    & No              & No                            & No                            & No                            & No                                      \\
    Turbulent diffusion       & No              & Eq. \eqref{eq:dturb_montreal} & Eq. \eqref{eq:dturb_montreal} & Eq. \eqref{eq:dturb_montreal} & Eq. \eqref{eq:dturb_montreal}           \\
    Rotation                  & No              & No                            & No                            & TZ97                          & TZ97                                    \\
    Loss of angular momentum
    by magnetized winds       & No              & No                            & No                            & M15                           & M15                                     \\
    Atmosphere                & Hopf            & VAL-C                         & VAL-C                         & VAL-C                         & VAL-C                                   \\
    \hline
\end{tabular}
        \end{center}
        Notes: $\alpha_{\rm ov, e}$ is the overshoot coefficient below the convective envelope, $T_{\rm cz}$ is the acoustic travel time inside the convective envelope. MP93 stands for \citealt{Michaud1993}. TZ97 for \citet{Talon1997b} and M15 for \citealt{Matt2015}.
    \end{table*}}

    {\renewcommand\arraystretch{1.75}
    \begin{table}[t]
        \caption{Values of observational constraints used to produce each calibrated solar models.}
        \label{table:snssm_obs}
        \begin{center}
        \begin{tabular}{L {0.14\textwidth} l l}
            \hline\hline
            Name                                          & Notation           & Value [cgs]                         \\
            \hline
            Age                                           & Age                & $4570$ [Myrs]                       \\
            Effective temperature                         & $\teff$            & $5772\pm1$ (1)                      \\
            Luminosity                                    & $L$                & $3.828\cdot 10^{33}\pm10^{28}$ (1)  \\
            Surface metal to hydrogen ratio               & $(Z/X)_{\rm s}$    & $0.0187\pm9\cdot10^{-4}$ (2)        \\
            Acoustic travel-time                          & $T_{\rm cz}$       & $1422\pm 20$ (3,4)                  \\
            Logarithm of surface ${}^7{\rm Li}$ abundance & $\log \epsilon_{\rm Li,s}$ & $0.96\pm0.06$ (2)           \\
            Surface angular velocity                      & $\Omega_\ss$       & $2.86\pm0.01$ (5)                   \\
            \hline
        \end{tabular}
        \end{center}
        Notes: (1) \citet{IAU2016}; (2) \citet{Asplund2021}; $\log\epsilon_{\rm Li,s} = \log(N_{\rm Li}/N_{\rm H}) + 12$; (3) \citet{Roxburgh2009,Deal2023}; (5) Solar surface angular velocity varies with latitude and moment of the solar cycle \citep[e.g. ][]{Beck2000}. The value here is chosen within a range of usual values of equatorial $\Omega_{\ss, \odot}$.
    \end{table}}

    {\renewcommand\arraystretch{1.75}
    \begin{table*}[t]
        \caption{Values of tunable parameters obtained after the calibration of each solar models.}
        \label{table:snssm_par}
        \begin{center}
        \begin{tabular}{l c c c c c}
            \hline\hline
            Name                  & SSM1       & SSM2      & NSSM1              & NSSM2              & NSSM3              \\
            \hline
            $\amlt$               & $1.640 $ & $2.026  $   & $1.971$            & $1.984$            & $1.971$            \\
            $Y_0$                 & $0.2549$ & $0.2626 $   & $0.2549$           & $0.2564$           & $0.2549$           \\
            $(Z/X)_0$             & \none    & $0.0212$    & $0.0192$           & $0.0196$           & $0.192$            \\
            $\alpha_{\rm ov, e}$  & \none    & $0.1408 $   & \none              & \none              & \none              \\
            $T_{\rm mix}$ [K]     & \none    & \none       & $2.596\cdot10^6$   & \none              & $2.633\cdot10^6$   \\
            $T_0$ [erg]           & \none    & \none       & \none              & $2.21\cdot10^{31}$ & $1.99\cdot10^{31}$ \\
            \hline
        \end{tabular}
        \end{center}
        Notes: We do not report here the uncertainties resulting from the Levenberg-Marquardt fit, as this method is known to underestimate such uncertainties. A better estimate would come from the usage of MCMC or Bayesian methods.
    \end{table*}}

    \paragraph{\textit{SSM1.}} The first standard solar model (SSM1) presented here is the model obtained with the default physics and numerical settings of \cesamxx (\texttt{'co'}). It includes the physical ingredients described in Sect. \ref{paragraph:comphys}, with no diffusion of chemical elements, nor any transport of angular momentum. Its initial Helium abundance, $Y_0$, and parameter $\amlt$ of MLT are calibrated so that it reproduces the solar $\teff$ and luminosity.

    \paragraph{\textit{SSM2.}} The second standard solar model (SSM2) is computed with the choice of input physics made to compute the first generation grid of stellar models for the PLATO mission, except that the convection is modelled with a traditional MLT with fixed $\amlt$, and is not entropy-calibrated. The numerical precision is higher (\texttt{'pl'}) and gravitational settling as well as turbulent mixing of chemical elements are included. The turbulent diffusion coefficient follows Eq. \eqref{eq:dturb_montreal}, with $\omega = 10^4$, $n = 4$ and the reference depth is set at a mass of $5\cdot 10^4$ below the surface. This choice is more suitable for F-type stars, and induces little mixing in the Sun. The treatment of turbulent diffusion shall be improved for the second generation of the grid. Contrary to SSM1, the atmosphere is here reconstructed with $\ttau$ relation of \citet{Vernazza1981}, which was calibrated to reproduce solar temperature gradient in the photosphere. To the previous choice of constraints and tunable parameters, we also tuned the initial $Z/X$ ratio and the overshoot parameter below the convective envelope to reproduce present solar $Z/X$ at the surface, and the acoustic travel time in the convective envelope.

    \paragraph{\textit{NSSM1.}} In a first non-standard model (NSSM1), we replace additional turbulent mixing adapted to F-type stars to a calibrated one. This mixing is still described by Eq. \eqref{eq:dturb_montreal}, but $\omega = 400$ and $n = 3$. The reference depth, in the present case located by a temperature $T_{\rm mix}$, is tuned, so that our model reproduces the present surface ${}^7{\rm Li}$ abundance of the Sun. We also kept  $Y_0$, $(Z/X)_0$, and $\amlt$ as tunable parameters and $T_{\rm eff}$, $L$, and $(Z/X)_{\rm s}$ as observational constraints. For all non-standard models here, the overshoot below the envelope is turned off and the atmosphere follows the $\ttau$ relation from \citet{Vernazza1981}.

    \paragraph{\textit{NSSM2.}} Then, NSSM2 has the same settings as SSM2 for the turbulent diffusion, however, rotation is taken into account following the \citet{Talon1997b} formalism as well as including the \citet{Matt2015}'s prescription for the loss of angular momentum by magnetized solar winds. This scaling relation depends on a multiplicative parameter $T_0 = 9.5 \times 10^{30}$ in the original paper. We took it as a free parameter and made it vary so that we recover the surface solar rotation age at present age. The period and lifetime of the disk are set to $10$ days and 5 Myrs, for lack of a better choice.

    \paragraph{\textit{NSSM3.}} Finally, the last non-standard model (NSSM3) combines the above two transport processes. These are controlled by $T_{\rm mix}$ and $T_0$.

    \subsection{Acoustic structure}

    \begin{figure*}[t]
        \begin{center}
            \includegraphics[draft=False, width=\textwidth]{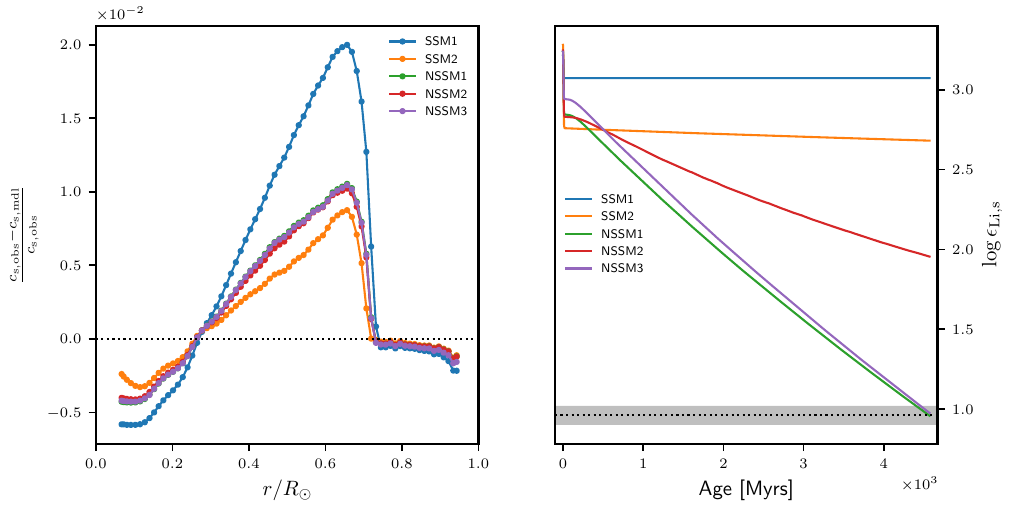}
            \caption{Left panel: Relative differences between inverted and modelled sound speed as a function of the radial coordinate, for SSM1 to NSSM3. The inverted sound speed is taken from \citep{Basu2000}. Right panel: Time evolution of the surface ${}^7{\rm Li}$ abundance for SSM1 to NSSM3. Horizontal black line is the present-day solar surface ${}^7{\rm Li}$ abundance as measured by \citet{Asplund2021}, and the shaded area represents the associated uncertainty.}
            \label{fig:sol_mdl}
        \end{center}
    \end{figure*}

    A classical test for assessing the accuracy of the structure of solar model is to compare the modelled sound speed $c_{\rm s, mdl}$ to the sound speed profile inverted using helioseismology $c_{\rm s, obs}$. For this reference profile, we used the data provided by \citet{Basu2000} inverted using frequency measurements of the Michelson-Doppler Imager (MDI) on-board Solar and Heliospheric Observatory (SoHO) satellite \citep{Scherrer1995}. The relative difference $c_{\rm s, mdl}$ and $c_{\rm s, obs}$ are displayed in Fig. \ref{fig:sol_mdl}, left panel. The highest relative differences reach $2\%$ for the simplest standard, SSM1, around $0.6R_\odot$. It should be stressed that the peak is higher, but comparable to what was shown in Fig. 1 of ML08 ($1.5\%$) with the previous version \cesamk. This model was computed with the solar chemical composition of \citet[AGS05]{Asplund2005} and the microscopic diffusion of MP93, while SSM1 used AAG21 and no diffusion. These different choices of modelling somewhat compensate each other: AGS05 is known to lo lead to inaccurate sound speed in the interior \citep[see e.g.][]{Asplund2009}, while accounting for the microscopic diffusion lower the error on the sound speed. The other three models (SSM2 and NSSM1-3) have display comparable error on their sound speed profiles. Their maximum error is still higher than what was shown in ML08. The reason is again that the solar abundance
determinations from \citet{Grevesse1993}  were used for the best models of ML08, which are known to lead to better agreement with inverted sound speed than when using AAG21 as it is the case for the models of the present paper \citep[see e.g.][]{Buldgen2019}.

    We note that a better agreement with the sound speed profile is not a sufficient criterion to make a better solar model \citep{Eggenberger2022,Buldgen2024,Buldgen2025}. Of all the models, the standard SSM2 model has the most accurate sound speed profile, while it does not incorporate the most complex physical ingredients. These models should be tested against other constraints than just their ability at reproducing the sound speed.

    \subsection{Lithium abundance}

    In Fig. \ref{fig:sol_mdl}, we display on the right panel the evolution of the surface ${}^7{\rm Li}$ abundance, as well as the targeted value. It is clear that we can no longer be satisfied with the standard model, even when it includes microscopic diffusion as SSM2. Indeed, almost no ${}^7{\rm Li}$ is depleted from the formation of the star to the present age of the Sun. The model NSSM2 that includes rotation, but no additional turbulent transport depletes more ${}^7{\rm Li}$ thanks to the enhanced transport of chemical elements brought by the shear-induced turbulence and the stronger coupling between the radiative zone and the convective envelope. However, the only way to achieve the present-day surface solar ${}^7{\rm Li}$ abundance is by adding an extra mechanism for the transport of the chemical elements. In \cesamxx, this can be done though an ad hoc turbulent mixing as it is the case in SSM1 and SSM3, but it could also be done by adding new processes like waves of (M)HD instabilities, similarly to models present in the literature \citep[e.g.][]{Talon2003,Fuller2015,Eggenberger2019a,Eggenberger2019b}.

    \section{Conclusion}

    In this paper, we present the main aspects of the numerical machinery,  covering the most recent developments regarding the physical processes that \cesamxx has the capacity to include and that were not available in earlier versions of \cesam or \cestam. We have shown, through a series of standard and non-standard solar models, the capabilities of \cesamxx for reproducing the acoustic structure of the Sun or the present-day surface solar ${}^7{\rm Li}$ abundance. Currently,  progress is also being made in the modelling of (M)HD instabilities, carrying angular momentum or chemical elements, in terms of the impact of magnetic fields on the stellar structure and evolution, on the reconstruction of atmospheres, and on other features.

    Recently, part of this overall effort has opened \cesamxx\ up to the community. We built a website\footnote{\url{https://www.ias.u-psud.fr/cesam2k20}} that puts togethr a 'quick start' guide, an improved documentation that is progressively being translated into English. On this website, we provide some models, in particular, all the models presented in this work, along with a link to the public repository. The public version will be updated regularly, with new published developments and bug fixing. The code is licensed under GPLv3, which means that the code can be changed, shared, or used for commercial use, as long as the source is made available with the same licence.

    \begin{acknowledgements}
        The authors warmly thanks everybody who played a role, even the tiniest, in the development of \cesamxx and earlier versions over the years, in particular Pierre Morel. The authors also thank A. Oreshina, V. Baturin, and all the people behind the SAHA-S EoS for their help in implementing it in \cesamxx. L. M. acknowledge support from the Agence Nationale de la Recherche (ANR) grant ANR-21-CE31-0018 and from the Max Planck Society (MPG) under project "Preparations for PLATO Science". M. D. acknowledge the support from CNES, focused on PLATO.
    \end{acknowledgements}

    \bibliographystyle{aa}
    \bibliography{biblio}

\end{document}